\documentclass[pre,aps,twocolumn]{revtex4-1}
\usepackage[latin1]{inputenc}
\usepackage[english]{babel}
\usepackage{graphicx}
\usepackage{color}
\usepackage{amsmath}
\begin{document}

\title{\bf\Large A Langevin canonical approach to the dynamics of
chiral systems. Populations and coherences.}

\author{H. C. Pe\~nate-Rodr\'{\i}guez $^{a,b}$
A. Dorta-Urra$^{c}$, P. Bargue\~no $^{*,d}$, G.
Rojas-Lorenzo$^{a,b}$ and S. Miret-Art\'es$^{**,b}$}
\affiliation{$^{a}$ Instituto Superior de Tecnolog\'{\i}a y
Ciencias Aplicadas, Ave. Salvador Allende y Luaces, Quinta de Los
Molinos, Plaza, La Habana 10600, Cuba
\\
$^{b}$ Instituto de F\'{\i}sica Fundamental (CSIC), Serrano 123
{\it E-28006, Madrid, Spain, ($^{**}$s.miret@iff.csic.es) }
\\
$^{c}$ Unidad asociada UAM-CSIC, Instituto de F\'{\i}sica
Fundamental (CSIC), Serrano 123 {\it E-28006, Madrid, Spain}
\\
$^{d}$ Departamento de F\'{\i}sica de Materiales, Universidad
Complutense de Madrid, {\it E-28040}, Madrid, Spain
($^*$p.bargueno@fis.ucm.es) }

\date{\today}

\begin{abstract}
A canonical framework for chiral two--level systems coupled to a
bath of harmonic oscillators is developed to extract, from a
stochastic dynamics, the thermodynamic equilibrium values of both
the population difference and coherences. The incoherent and
coherent tunneling regimes are analyzed for an Ohmic environment
in terms of a critical temperature defined by the maximum of the
heat capacity. The corresponding numerical results issued from
solving a non-linear coupled system are fitted to approximate
path--integral analytical expressions beyond the so-called
non-interacting blip approximation in order to determine the
different time scales governing both regimes.
\end{abstract}
\maketitle

\section{Introduction}
The description of many phenomena in terms of a two level system
(TLS) can be found in different areas of chemical physics (chiral
molecules
\cite{Harris1978,Quack1986,MacDermott2004,MacDermott2012},
electron transfer reactions \cite{Weiss1999,Transferbook}),
quantum optics \cite{Scullybook}, high energy physics
\cite{Rosner2001,capo1,capo2} and quantum computation
\cite{Ladd2010}, among many others fields of research. However,
the isolated TLS is not complex enough to provide a richer and
more complete image of the basic microscopic processes underlying
its dynamics. For example, very often the system is coupled to a
more extended system or environment consisting of many degrees of
freedom usually represented by an infinite set of harmonic
oscillators. Thus, several extensions of the isolated TLS have
been taken into account along the years to account properly for
dissipative and stochastic processes due to interactions with an
environment at finite temperature. In general, the TLS is linearly
coupled to the coordinates of a bath of noninteracting
oscillators, whose properties are encoded in their spectral
density \cite{Leggett1987} (the spin--boson model). Several
approaches within the matrix density formalism have been proposed
and implemented to determine the time evolution of the dissipative
TLS such as  path--integral methods or the Bloch--Redfield
formalism
\cite{Redfield1957,Senitzki1963,Harris1981,Harris1983,Hartmann2000,Vincenzo2005,Isabel2011}.
In the path--integral formalism, the so--called non-interacting
blip approximation (NIBA) breaks down for non-zero bias and low
temperatures even for weak damping, and a theory beyond the NIBA
for a biased system has been developed \cite{Weiss1999}.
Variational calculations have also been carried out for both the
symmetric (or unbiased) \cite{Silbey1984} and asymmetric (or
biased) cases \cite{Harris1985}. Alternatively, following an
approach to the dynamics of the TLS due to Feynman in his elegant
dynamical theory of the Josephson effect \cite{Feynman}, it was
shown that classical and quantum mechanics may be embedded in the
same Hamiltonian formulation by using complex canonical
coordinates \cite{Strocchi1966,Heslot1985}. In particular, for
electronically non--adiabatic processes, Meyer and Miller
\cite{Meyer1979} (following some earlier work by McCurdy and
Miller \cite{McCurdy1978}) shown the possibility of representing
each electronic state by a pair of classical action-angle
variables. However, instead of these variables, cartesian
electronic variables have been extensively used because of the
simpler algebraic form of the Hamiltonian. This approach and some
of its extensions have been employed by several groups (for a
complete list of references on the development and applications of
this method, see \cite{Miller2009} and references therein).
Furthermore, by using the action--angle variables, this formalism
has also been recently implemented to study the dissipative
dynamics (zero temperature) of chiral molecules. In this case, the
asymmetry of the assumed double well potential model is due to the
parity violating energy difference
\cite{Bargueno2011,Penate2012,Dorta2012}. A natural extension of
this dynamics is to include the temperature effect along with a
noise term.

In this spirit, this is the first of a series of upcoming articles
in which we describe, implement and apply a Langevin canonical
approach to the dynamics of the chiral TLS interacting with an
environment at finite temperatures.
%When interactions between the
%TLS and the environment are switched on, the thermal expectation
%values are obtained when the asymptotic time limit of the average
%of stochastic trajectories is reached.
Section II is devoted to introduce the canonical formalism for an
isolated chiral TLS, as well as its usual connection with the
density matrix approach and the corresponding thermodynamics.
Finite temperature effects are included by means of noise--induced
dynamics via a Caldeira--Leggett--like Hamiltonian. Numerical
simulations of this stochastic approach are presented and
discussed in Section III, where population differences (which are
proportional, for example, to the optical activity in chiral
systems \cite{MacDermott2004,MacDermott2012})
as well as coherences are issued by assuming an Ohmic
regime in a broad range of temperatures. In particular, the
competition between the tunneling and the asymmetry or bias
displayed by the TLS is analyzed in terms of a critical
temperature which is defined by the maximum of the heat capacity
and separates the coherent and incoherent tunneling regimes. In
all cases, these two regimes are discussed in the framework of a
theory beyond the NIBA, path--integral analytical expressions in
order to validate this stochastic dynamics. Even more, the
corresponding thermodynamic functions are obtained from this
stochastic analysis at asymptotic times where the thermal
equilibrium is reached. Finally, Section IV presents the main
conclusions and future perspectives of the formalism here
employed.

\section{Theory}

\subsection{A canonical formalism for chiral two level systems}

Let us consider an isolated chiral TLS described by the
Hamiltonian $\hat H=\delta \hat\sigma_{x}+\epsilon \hat\sigma_{z}$
where $\sigma_{x,z}$ stand for the Pauli matrices. The isolated
system is usually modelled in a phenomenological way by a
two-well (asymmetric) potential within the Born-Oppenheimer
approximation. From the knowledge of the eigenstates,
$|1\rangle,|2\rangle$, the left and right states (or chiral
states), $|L\rangle$ and $|R\rangle$, respectively, can be
expressed by means of a rotation of angle $\theta$ given by $\tan
2\theta = \delta/\epsilon$, where $\langle L | \hat H |R \rangle
=- \delta$ (with $\delta > 0$) describes the tunneling rate  and
$2\epsilon=\langle L|\hat H|L\rangle -\langle R|\hat H|R\rangle $
($\epsilon$ can be positive or negative) accounts for the
asymmetry due to the electroweak parity violation (for a chiral
system) or any other bias term (for example, a magnetic field).

Among other interesting representations of the isolated system
\cite{Quantumbook}, an alternative and useful way of looking at it
is based on the polar decomposition of the complex amplitudes
entering the wave function. The solutions of the time-dependent
Sch\"odinger equation ($\hbar=1$)
\begin{equation}
i\,\partial_{t}|\Psi(t)\rangle=\hat H|\Psi(t)\rangle
\label{SE}
\end{equation}
can be written as
$|\Psi(t)\rangle=a_{L}(t)|L\rangle+a_{R}(t)|R\rangle$. If the
complex amplitudes are written in polar form as
$a_{L,R}(t)=|a_{L,R}(t)|e^{i\Phi_{L,R}(t)}$, and the population
and phase differences between chiral states are defined as
$z(t)\equiv|a_{R}(t)|^{2}-|a_{L}(t)|^{2}$ and
$\Phi(t)\equiv\Phi_{R}(t)-\Phi_{L}(t)$, respectively, it is an
easy exercise to prove that the average energy in the normalized
$|\Psi(t)\rangle$ state is given by $\langle \Psi|\hat
H|\Psi\rangle= -2\delta\sqrt{1-z^{2}}\cos\Phi+2\epsilon z \equiv
H_{0}$, where $H_{0}$ represents a Hamiltonian function. As $z$
and $\Phi$ can be seen as a pair of canonically conjugate
variables, the Heisenberg equations of motion (which are formally
identical to the Hamilton equations) are easily derived from $\dot
{z} = -
\partial H_{0} /
\partial \Phi$ and $\dot {\Phi} = \partial H_{0} / \partial z$.
Explicitly, the non-linear coupled equations describing the
isolated system in these canonical variables are
\begin{eqnarray}
\label{TLQSeq}
\dot z &=&-2\delta\sqrt{1-z^2}\sin \Phi \nonumber
\\
\dot \Phi &=&2\delta\frac{z}{\sqrt{1-z^2}}\cos \Phi+
2\epsilon.
\end{eqnarray}
Thus, Eqs. (\ref{TLQSeq}) are totally equivalent to the usual time
-dependent Schr\"odinger equation (\ref{SE}). In fact, the exact
quantum beating expression can be obtained by noting that $H_{0}$
is a conserved magnitude \cite{Bargueno2011}. For simplicity, the
adimensional time $t\rightarrow 2\delta\, t$ will be used in the
rest of this work. This re-scaling implies that the Hamiltonian
function $H_{0}$ is expressed again as
\begin{equation}
H_{0}=-\sqrt{1-z^{2}}\cos\Phi+\frac{\epsilon}{\delta} z. \label{H}
\end{equation}
Notice that the first term of the Hamiltonian function (\ref{H})
accounts for the tunneling process and the second one for the
underlying asymmetry (due to a bias or the parity-violating energy
difference) putting on evidence the two competing processes  in
this simple dynamics.

Finally, the connection between the canonical and the density
matrix formalism can be established as follows. Let $\hat\rho$ be
the density matrix for the TLS, whose matrix elements are given by
$\rho_{R,R}=|a_{R}|^{2}$, $\rho_{L,L}=|a_{L}|^{2}$,
$\rho_{L,R}=a_{L}a_{R}^{*}$ and $\rho_{R,L}=a_{R}a_{L}^{*}$. On
the other hand, from $\langle \Psi(t)|\hat
H|\Psi(t)\rangle=\mathrm{Tr}(\hat\rho \hat H )=H_{0}$, one obtains
that the time average values of the Pauli operators are given by
\begin{eqnarray}
\label{correspondence} \langle \hat \sigma_{z}\rangle_t&=&
\rho_{R,R}-\rho_{L,L} = z  \nonumber \\
\langle \hat \sigma_{x}\rangle_t&=& \rho_{R,L}+\rho_{L,R} =
-\sqrt{1-z^{2}}\cos\Phi \nonumber \\
\langle \hat \sigma_{y}\rangle_t&=& i \rho_{R,L} - i \rho_{L,R}
=\sqrt{1-z^{2}}\sin\Phi,
\end{eqnarray}
which is consistent with $\langle\hat H\rangle=\delta\langle \hat
\sigma_{x}\rangle+ \epsilon \langle \hat \sigma_{z}\rangle$ and
\begin{equation} \label{nodamping}
\langle \hat \sigma_{x}\rangle_t^2 + \langle \hat
\sigma_{y}\rangle_t^2 + \langle \hat \sigma_{z}\rangle_t^2 = 1 .
\end{equation}
%

%In addition, following \cite{Weiss1999}, one also finds $\langle
%\hat \sigma_{y}\rangle=- d \langle \hat \sigma_{z}\rangle / dt$.

The time population difference can also be split into two
components which are symmetric and antisymmetric under the
inversion operation consisting of replacing $\epsilon$ by $-
\epsilon$.

\subsection{Stochastic dynamics of chiral systems}

The dynamics of the isolated chiral TLS can be reduced to simply
solve Eqs. (\ref{TLQSeq}) and then form appropriate combinations
of $z$ and $\Phi$ to recover the populations and coherences. When
dealing with interactions with the environment consisting of a
high number of degrees of freedom, more sophisticated theoretical
approaches are needed. They can be widely classified into three
frameworks according to the picture of quantum mechanics used
\cite{Weiss1999,Petruccione2006}: the density operator formalism
and the stochastic Schr\"odinger equation (Schr\"odinger and
interaction picture), and the generalized Langevin equation
(Heisenberg picture). Due to the fact the formalism used here to
describe the dynamics of a TLS is very much attached to the
definition of a Hamiltonian function (\ref{H}), the last framework
is much more convenient, apart from being much less employed in
the theory of open quantum systems. Within this canonical
formalism, a Caldeira--Leggett--like Hamiltonian,
\cite{Leggett1987} where a bilinear coupling between the TLS and
the environment is assumed, is usually found in the literature. In
particular, we have recently developed this formalism to study the
dissipative dynamics of chiral systems
\cite{Bargueno2011,Penate2012,Dorta2012}.

As previously stated \cite{Bargueno2011}, noting that $\Phi$ and
$z$ play the role of a generalized coordinate and momentum,
respectively, one can introduce interactions with the environment
by means of a system-bath bilinear coupling {\it via} a
Caldeira--Leggett--like Hamiltonian expressed as
\begin{eqnarray}
\nonumber H &=& H_{0}+ \frac{1}{2}\sum_{i}\left(\Lambda_{i}p^{2}_{i}+\frac{x^{2}_{i}\omega_{i}^{2}}
{\Lambda_{i}}\right)\\
&-&\Phi \sum_{i}c_{i}x_{i}+\Phi^{2}\sum_{i}c^{2}_{i}\Lambda_{i},
\end{eqnarray}
where the sums run over the  coordinates of the bath oscillators
$\{p_i,x_i \}$ and $\Lambda_{i}$, $c_i$ and $\omega_{i}$ are
suitable dimensionless constants representing generalized masses,
couplings with the environment, and oscillator frequencies,
respectively. Although the requirement of a bilinear coupling has
been relaxed in a previous publication \cite{Dorta2012}, it will
be retained here for simplicity.

It should also be noticed that the usual spin--boson Hamiltonian
has been implemented in the Meyer--Miller--Stock--Thoss
representation by coupling the bath position coordinate with the
population difference of the TLS (see Eq. (2.11) of
\cite{Wang1999}). On the contrary, in the approach here employed,
it is the phase difference of the TLS the canonical variable which
is coupled to the bath position coordinate, not the population
difference. It follows quite closely that employed in the field of
condensed matter, the dynamics of a Josephson junction
\cite{Weiss1999}. This phase difference is coupled to the degrees
of freedom of the bath which also acts as a source of phase
fluctuations.
%In any case, it is expected that thermal equilibrium
%is reached independent on the coupling scheme assumed.
%Moreover, the stochastic
%approach here presented reproduces fairly good path--integral NIBA
%calculations for both the incoherent and coherent tunneling
%regimes, as it will be shown.

Within this scheme, the corresponding coupled Langevin-type
dynamical equations are given by
\begin{eqnarray}
\label{stochohm}
\dot z &=&-\sqrt{1-z^2}\sin \Phi \nonumber
\\
&-&\int^{t}_{0}\gamma(t-t') \Phi(t') \dot \Phi(t') \, dt' + \xi(t)
\nonumber \\
\dot \Phi &=&\frac{z}{\sqrt{1-z^2}}\cos \Phi+
\frac{\epsilon}{\delta} ,
\end{eqnarray}
where the time-dependent friction (damping kernel) is expressed as
\begin{equation}
\gamma(t)=\sum_{i}\Lambda_{i}c_{i}^{2}\cos \omega_{i}(t-t')
\end{equation}
and the fluctuation force or noise is given by
\begin{eqnarray}
\xi(t)=&\sum_{i}&c_{i}\Lambda_{i}\left(x_{i}(0)\cos \omega_{i}
t + p_{i}(0)\sin \omega_{i} t\right) \nonumber \\
&-&c_{i}^{2}\Lambda_{i}\Phi(0) \cos \omega_{i} t
\end{eqnarray}
which depends on the initial conditions of both the system and the
bath. Taking the bath oscillators as classical variables
(classical noise), moderate--to--high temperature regimes are
expected to be properly described by this approach, as will be
discussed through the rest of the manuscript.

If a Markovian regime is assumed, the standard properties of the
fluctuation force (Gaussian white noise) are given by the
following canonical thermal averages: $\langle
\xi(t)\rangle_{\beta}=0$ (zero average) and $\langle
\xi(0)\xi(t)\rangle_{\beta}=m k_{B}T\gamma \delta (t)$
(delta-correlated) where $\beta = (k_B T)^{-1}$, $k_B$ being
Boltzmann's constant. The friction is then described by
$\gamma(t)=2\gamma\delta(t)$, where $\gamma$ is a constant and
$\delta(t)$ is Dirac's $\delta$--function (not to be confused with
the $\delta$-parameter describing the tunneling rate). Thus, in
this regime, Eqs. (\ref{stochohm}) read now
\begin{eqnarray}
\label{stochohm1} \dot z &=&-\sqrt{1-z^2}\sin \Phi - \gamma \dot
\Phi(t)  + \xi(t) \nonumber \\
\dot \Phi &=&\frac{z}{\sqrt{1-z^2}}\cos \Phi+
\frac{\epsilon}{\delta} .
\end{eqnarray}
The corresponding solutions provide stochastic  trajectories for
the population and phase differences encoding all the information
on the dynamics of the non-isolated TLS. These solutions  are
dependent on the four dimensional parameter space
$(\epsilon,\delta,\gamma,T)$, apart from the initial conditions
$z(0) = z_0$ and $\Phi (0) = \Phi_0$. In terms of the averages of
Pauli operators, we have now the condition
\begin{equation} \label{damping}
\langle \hat \sigma_{x}\rangle_t^2 + \langle \hat
\sigma_{y}\rangle_t^2 + \langle \hat \sigma_{z}\rangle_t^2 < 1 .
\end{equation}
instead of condition (\ref{nodamping}).

The use of classical noise imposes some restrictions on the range
of temperatures where this approach remains valid. At high
temperatures, $\beta^{-1} \gg \hbar \gamma$ (or $\gamma^{-1} \gg
\hbar \beta$) thermal effects are going to be predominant over
quantum effects which become relevant, in general, at times of the
order of or less than $\hbar \beta$, sometimes also called thermal
time. However, in general, at very low temperatures, $\beta^{-1}
\ll \hbar \gamma$ (or $\gamma^{-1} \ll \hbar \beta$) the noise is
colored and its correlation function is complex and our approach
is no longer valid. The dissipative dynamics in the classical
noise regime is obtained at zero  noise or zero temperature
\cite{Sanzbook}. This regime has already been considered elsewhere
\cite{Dorta2012}.

Canonical thermal averages of population and phase differences as
a function of time $\langle z(t) \rangle_{\beta}$ and $\langle
\Phi (t) \rangle_{\beta}$ are issued from running a high number of
stochastic trajectories. The role of initial conditions has been
extensively discussed in the literature (see, for example,
\cite{Weiss1999,Petruccione2006}) and we are not going to deal
with this important issue. In our dynamical study, the system will
be prepared in one of the chiral states, left or right ($z_0=
0.999$ or $-0.999$ in order to avoid initial singularities), and
the initial phase difference $\Phi_0$ will be uniformly
distributed around the interval $[- 2 \pi, 2 \pi]$. This approach
should recover the main equilibrium thermodynamics properties of
the non-isolated TLS from the stochastic dynamics at asymptotic
times.

\subsection{Thermodynamics of chiral two-level systems}

A detailed analysis of the thermodynamics of non-interacting
chiral molecules assuming a canonical distribution was carried out
elsewhere \cite{Bargueno2009}. In particular, thermal averages of
pseudoscalar operators were extensively analyzed. The canonical
thermal average of an observable $X$ is defined as $\langle X
\rangle_{\beta}= \mathrm{Tr}(\rho_{\beta}X)$ where
$\rho_{\beta}=Z^{-1}e^{-\beta H_0}$, $H_0$ is given by (\ref{H})
and $Z$ is the partition function. For chiral states,
$Z=2\cosh(\beta \Delta)$ with
$\Delta=\sqrt{\delta^{2}+\epsilon^{2}}$ and the corresponding
averages for the population difference and coherences (in the L--R
basis) are then calculated to give
\begin{eqnarray}\label{thermo}
\langle z \rangle_{\beta} \equiv \langle \hat
\sigma_{z}\rangle_{\beta}&=&\frac{\epsilon}{\Delta}\tanh(\beta
\Delta) \nonumber
\\
\langle \hat \sigma_{x}\rangle_{\beta}&=&\frac{\delta}{\Delta}\tanh(\beta \Delta)
%\nonumber
%\\
%\langle \hat
%\sigma_{y}\rangle_{\beta}&=&\frac{1}{\Delta}\frac{d}{dt}\langle
%\hat \sigma_{z}\rangle_{\beta}.
\end{eqnarray}

From the knowledge of the partition function, the remaining
equilibrium thermodynamical functions are also easily deduced such
as the Helmholtz free energy, the entropy, the heat capacity, etc.
From such an analysis, a critical temperature given by
\cite{Bargueno2009}
\begin{equation}
T_c \sim \frac{\Delta}{k_B 1.2}
\end{equation}
is derived when $\langle z \rangle_{\beta}$ displays an inflection
point and the heat capacity a maximum as a function of the
temperature. At temperatures higher than $T_c$, the effect of
$\epsilon$ is masked by thermal effects which tend to wash out the
population difference $z$ (racemization). At temperatures lower
than $T_c$, the value of the ratio $\epsilon / \delta$ is
critical. When this ratio is close to unity, $\langle z
\rangle_{\beta}$ is determined by the competition between
tunneling and asymmetry or bias. When it is much greater than one,
the tunneling process plays a minor role and $\langle z
\rangle_{\beta}$ keeps more or less its initial value. Finally,
when this ratio is much less than one, the racemization is always
present.

In a previous work \cite{Bargueno-pccp}, it was showed that at
very low temperatures, a chiral or two level bosonic system could
display condensation as well as a discontinuity in the heat
capacity (reduced temperatures $k_B T / \Delta \leq 1$). In this
work, we are going to assume that in this temperature regime the
Maxwell-Boltzman (canonical) distribution still applies, instead
of the Bose-Einstein statistics, in order to know if the
thermodynamics can also be reached within this formalism with a
classical noise. Otherwise, a colored noise and a generalized
Langevin equation formalism should be applied.

Finally, it is worth stressing that the thermodynamic functions
are independent on the friction coefficient in the weak coupling
limit. Thus, our average values of population differences issued
from this stochastic dynamics are independent on the friction
coefficient as time goes to infinity, that is, when the thermal
equilibrium with the bath is reached. In the strong coupling
limit, this fact no longer holds \cite{Ingold2009}.

\section{Results}

\subsection{Numerical details}

The general strategy consist of solving the pair of non-linear
coupled equations (\ref{stochohm1}) for the canonical variables
under the action of a Gaussian white noise, which is implemented
by using an Ermak--like approach \cite{Ermak1980,Allenbook}.
%In essence, Ermak's method
%is based on integrating the equations of motion over a time
%integral, under the assumption that the stochastic forces remain
%approximately constant \cite{Allenbook}. We note that, contrary to
%the commonly employed Ermak--like algorithms, only a random
%variable is included instead of two.
Notice that in the Langevin--like coupled equations to be solved,
the noise term only appears in the equation of motion of the
$z$--variable. The phase variable has been taken to be uniformly
distributed between $-2\pi$ and $2\pi$. In order to avoid the
singular behavior found at $z\rightarrow \pm 1$, the initial value
for $z$ will be chosen to be around $\pm 0.999$, very far from the
equilibrium condition. Moreover, when running trajectories there
are some of them visiting "un-physical" regions, that is, $|z|>1$.
This drawback is mainly associated with the intensity of the noise
since, for large values of it (which depends on both the
temperature and the friction coefficient), the stochastic
$z$--trajectories can become unbounded. To overcome this problem,
we have implemented a {\it reflecting} condition such that, for
instance, when the trajectory reaches $z>1$, we change its value
to $2-z$. The time step used goes from $10^{-6}$ to $10^{-4}$
(dimensionless units) depending strongly on the regime to be
explored in order to follow properly the corresponding dynamics:
high or moderate temperature regimes and localized or delocalized
regimes. As noted in \cite{Bargueno2011}, unstable trajectories
can also be found for certain values of $\epsilon$, $\delta$ and
$\gamma$ in the simple case of dissipative but non--noisy
dynamics. As this problem persists in case of dealing with
stochastic trajectories, not every triple
$(\epsilon,\delta,\gamma)$ gives place to a stable trajectory. In
these cases, the time evolution of individual trajectories is not
possible and a previous stability analysis is mandatory. However,
in the stable case, a satisfactory description of population
differences and coherences has been achieved by running up to $
10^{4}$ trajectories in all cases considered.

%The role of the number of valid trajectories employed
%to obtained average magnitudes is shown in Fig. (\ref{fig1}) for
%the case of $T=10$ (in units of $\Delta$), $\epsilon=\delta=1$, $\gamma=0.1$ and a timestep of $10^{-3}$.
%An increasing smoothness of the average is observed for larger values of $N$, as expected.

%\begin{figure}%[!h]
%\includegraphics[width=0.30\textwidth,height=0.40\textwidth,angle=-90]
%{trayectorias.epsi} \caption{\label{fig1}Time dependence of $\langle z \rangle_{\beta}$ for
%different number of valid trajectories, $N$. $N=10,10^{2},10^{3}$ and $10^{4}$ for panels $a$--$d$.
%See text for details.}
%\end{figure}

\subsection{Population differences and coherences}

In this subsection, thermal effects both in the population
difference and coherences will be studied, extending previous
results in which only dissipative dynamics (that is, at zero
temperature) was considered \cite{Dorta2012}. Specifically, we
will focus on the role of the critical temperature such that,
roughly speaking, separates coherent from incoherent tunneling.
Regarding the internal dynamics of the TLS, we start analyzing the
delocalized regime where $\delta > \epsilon$. In particular, we
have taken $\delta=1$ and $\epsilon=0.5$. Given these values, the
critical temperature is $T_{c}\sim 1$ in units of $\Delta$. Thus,
to study its effects in the dynamics, temperatures ranging from
200 to 0.4 (in units of $\Delta$) have been considered. As the
friction has been taken to be a constant value, $\gamma=0.1$, both
moderate and high temperature regimes ($k_{B}T\sim \hbar\gamma$
and $k_{B}T\gg \hbar\gamma$, respectively) are covered. The
propagation time step has been taken to be between $10^{-6}$ (high
temperatures) and $10^{-4}$ (moderate temperatures).

\begin{figure}[h!]
\includegraphics[width=0.5\textwidth,height=0.5\textwidth,angle=0]
{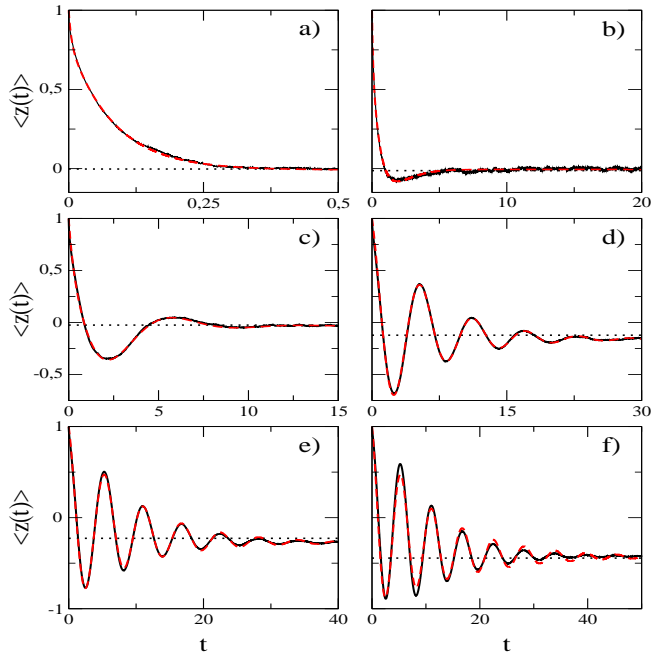} \caption{\label{fig1} Time dependence of $\langle z (t)
\rangle_{\beta}$ for different temperature values (in units of
$\Delta$): (a) 200, (b) 40, (c) 20, (d) 4, (e) 2 and (f) 0.4.
Black dashed curves are issued from solving Eqs.(\ref{stochohm1})
and red dashed curves from Eqs. (\ref{incoh}) (incoherent regime)
and (\ref{coh}) (coherent regime).}
\end{figure}

Trajectory averages (black dashed curves) of the population
differences, $\langle z (t) \rangle_{\beta}$, are plotted in Fig.
(\ref{fig1}) for the range of temperatures previously given above.
Left panels ($a$, $c$ and $e$) show calculations for $T = 200, 20$
and $2$, respectively. Right panels ($b$, $d$ and $f$) show the
calculations for $T = 40, 4$ and $0.4$. The system reaches the
thermal equilibrium value, given by the first expression of Eq.
(\ref{thermo}), at asymptotic times, which is plotted by a dotted
line in all the panels.
%The corresponding thermal times are given by $t \sim \hbar \beta$,
%shown in each panel due to the different values of the temperature.
Thus, as expected, more time is needed to reach the thermodynamic
equilibrium at lower temperatures since the coherent tunneling
(the oscillation regime between the two states) is dominating the
dynamics. The incoherent tunneling which prevails at the two
highest temperatures analyzed (panels a and b) leads to
racemization very rapidly.
%Specifically, the system gets equilibrated with the
%environment at $t=0.5, 10, 15, 30, 40$ and $50$, for $T=100, 20,
%10, 2, 1$ and $0.2$, respectively.
%
When the temperature approaches the critical temperature, $T_{c}$,
(panels d and e) the competition between the thermal and the
internal energy scale of the TLS (roughly given by $k_{B}T$ and
$\hbar \Delta$, respectively) gives place to the appearance of the
oscillating profile, fingerprint of coherent tunneling. When the
temperature is again lowered (panel f), the tunneling process
dominates the dynamics (remember that $\delta / \epsilon > 1$),
and more pronounced oscillations become more persistent in time.
Thus, for a fixed value of the friction coefficient, the critical
temperature roughly defines the coherent--incoherent transition.

In order to extract more physical information from our approach,
these stochastic calculations are fitted to analytical expressions
describing both coherent and incoherent tunneling. In the weak
Ohmic damping limit and moderate--to--high temperature regime the
path-integral results (beyond the NIBA framework) show a
dependence on time according to \cite{Weiss1999}
\begin{equation}\label{incoh}
\langle z (t) \rangle_{\beta} =\langle z \rangle_{\beta}+
\left[1-\langle z \rangle_{\beta}\right]e^{-\gamma_{2}t},
\end{equation}
for the incoherent regime and
\begin{eqnarray}\label{coh}
\langle z (t) \rangle_{\beta} &=&a_{1}e^{-\gamma_{1}t}+\langle z
\rangle_{\beta} \nonumber \\
&+&\left[(1-a_{1}-\langle z \rangle_{\beta})\cos \Omega
t+a_{2}\sin \Omega t \right]e^{-\gamma t},
\end{eqnarray}
for the coherent regime. In these two equations, $\gamma_2,
a_{1},a_{2},\gamma_{1},\gamma,\Omega$ are considered as free
parameters and $\langle z \rangle_{\beta}$ is given by the first
expression of Eq. (\ref{thermo}). Notice the fairly good fittings
of the average of the stochastic $z$--trajectories to Eqs.
(\ref{incoh}) and (\ref{coh}) (displayed by dashed red lines) for
all panels of Fig. \ref{fig1}. $\Omega$ can be interpreted as the
oscillation frequency of an effective damped harmonic oscillator
(Rabi-type) and $\gamma_1$, $\gamma_2$ and $\gamma$ the different
relaxation rates in this very involved dynamics. In particular,
$\gamma_2$ gives the effective decay rate for the incoherent
tunneling and $\gamma_1$ and $\gamma$ two effective decays for the
coherent tunneling with different weights. In other words, for
this last regime, $\Omega$ and $\gamma$ give us globally the two
different time scales observed in this stochastic dynamics since
$\gamma_1$ gives the incoherent contribution in this coherent
regime. All of these parameters are related by quite cumbersome
expressions according to the path-integral method. The quality of
the fitting should be related to the right behavior underlying by
the stochastic trajectories.

\begin{figure}[!h]
\includegraphics[width=0.5\textwidth,height=0.5\textwidth,angle=0]
{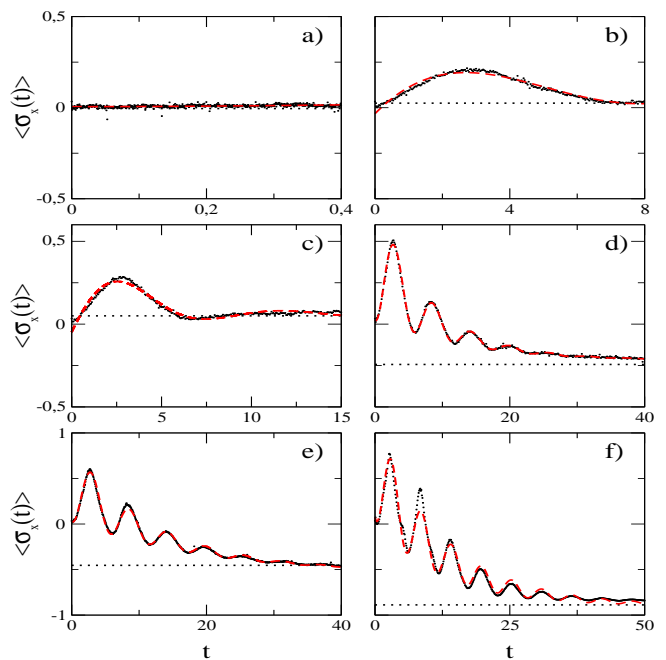} \caption{\label{fig2} Time dependence of the coherence
$\langle \hat \sigma_{x}\rangle_{\beta}$ for different temperature
values (in units of $\Delta$): (a) 200, (b) 40, (c) 20, (d) 4, (e)
2 and (f) 0.4. Black dashed curves are issued from solving
Eq.(\ref{stochohm1}) and red dashed curves from Eq. (\ref{coh2}).
For simplicity in the legend of the vertical axis, the hat of the
Pauli operator in the $x$-direction has been removed. }
\end{figure}

Let us consider now coherences. However, instead of analyzing the
average values of the phase difference given by $\langle \Phi (t)
\rangle_{\beta}$, we have calculated the thermal average value of
$\langle \hat \sigma_{x} (t) \rangle_{\beta}$ issued from our
stochastic trajectory analysis from the second expression of Eq.
(\ref{correspondence}). These values are plotted in Fig.
\ref{fig2} for the same temperature values as reported in Fig.
\ref{fig1} (the same temperature value is assigned to each panel).
Thus, we are considering $\langle \hat \sigma_x (t) \rangle =
\langle - \sqrt{1 - z^2} \cos \Phi \rangle$ and the stochastic
behavior on time is given by the black dashed curves, as before.
Similar comments with respect to the incoherent and coherent
regimes are also applicable here. Even more, in the weak Ohmic
damping limit and moderate--to--high temperature regime our
stochastic values are again fitted to the corresponding expression
provided by the path-integral method according to \cite{Weiss1999}
\begin{eqnarray}\label{coh2}
\langle \hat \sigma_{x}\rangle (t)&=&b_{1}e^{-\gamma_{1}t}+\langle \hat \sigma_{x}\rangle_{\beta}
\\
&+&\left[-(b_{1}+\langle \hat \sigma_{x}\rangle_{\beta})\cos \Omega t+b_{2}\sin \Omega t \right]e^{-\gamma t},
\nonumber
\end{eqnarray}
where $b_{1},b_{2}$ are again taken as free parameters and
$\langle \hat \sigma_{x}\rangle_{\beta}$ is given by the second
expression of Eqs. (\ref{thermo}). As before, the same
interpretations of $\Omega$, $\gamma_1$ and $\gamma$ are
pertinent.

The consistency of the fittings in Figs. \ref{fig1} and \ref{fig2}
is supported by the obtention of identical numerical values for
those parameters which are common to both expressions in Eqs.
(\ref{coh}) and (\ref{coh2}). This is illustrated in Table
\ref{table1} for only two temperatures, $T=4$ and $T=40$. The
different time scales clearly manifest in such a table.

\begin{table}
    \begin{tabular}{|c|c|c|c|c|c|}
        \hline
        ~                               & $T$ & $\epsilon/\delta$ & $\gamma_{1}$              & $\Omega$  & $\gamma$  \\ \hline
        $\langle\hat\sigma_{x}\rangle$  & 4 & 0.5               & 0.0164                    & -1.120    &  0.015    \\ \hline
        $\langle z \rangle$             & 4 & 0.5               & 0.0162                    & -1.120    & 0.015     \\ \hline
        $\langle\hat\sigma_{x}\rangle$  & 4 & 1.0               & 0.0210                    & -1.410    &  0.013    \\ \hline
        $\langle z \rangle$             & 4 & 1.0               & 0.0200                    & -1.410    &  0.012    \\ \hline
        $\langle\hat\sigma_{x}\rangle$  & 40  & 1.5               & 0.1220                    & -1.800    &     0.067 \\ \hline
        $\langle z \rangle$             & 40  & 1.5               & 0.1220                    & -1.798    & 0.072     \\ \hline
        $\langle\hat\sigma_{x}\rangle$  & 40  & 1.0               &   0.1060                  & -1.407    & 0.069     \\ \hline
        $\langle z \rangle$             & 40  & 1.0               & 0.1160                    & -1.408    &  0.076    \\ \hline
    \end{tabular}
\caption{Numerical fitting of the average of stochastic
trajectories corresponding to population differences and
coherences (Eqs. (\ref{coh}) and (\ref{coh2})). For simplicity,
only $T=4$ and $40$ and certain values of the ratio $\epsilon /
\delta$ are shown. All the quantities are dimensionless.}
\label{table1}
\end{table}

\subsection{Thermodynamics from stochastic dynamics}

Finally, it should be noticed that the formalism here presented
describes rather accurately the time dependence of both the
population differences and coherences, being an alternative way to
solve the time--dependent Schr\"odinger equation. It is also worth
noting the capabilities of this approach to describe properly
equilibrium thermodynamics in a large interval of temperatures,
ranging from moderate to high ones. In the very low-temperature
regime, where $k_{B}T\ll\hbar \gamma$, quantum noise effects are
expected to occur. In this regime, the non--commutativity of the
canonical variables describing the bath, $[x_{i},p_{i}]\ne 0$,
would lead to $[\xi(t),\xi(t')]\ne0$. Thus, within this range of
temperatures, the approximation here employed should break up.
However, in our study, we have assumed that the thermodynamical
behavior at low temperatures is also described by a canonical
distribution and correct values should also be obtained under this
assumption.

\begin{figure}
\includegraphics[width=0.45\textwidth,height=0.45\textwidth,angle=-90]
{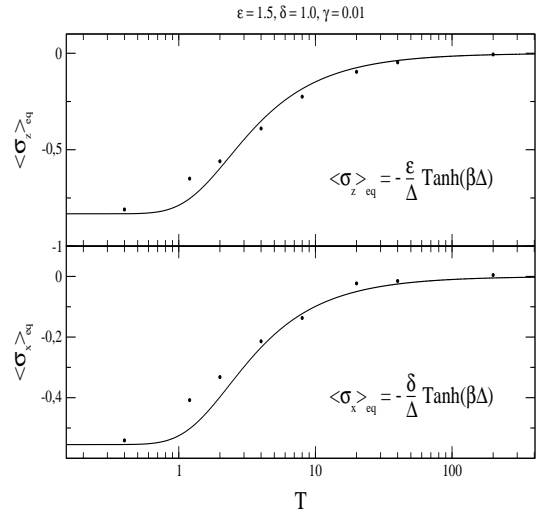} \caption{\label{fig3} Thermal dependence of the
population differences and coherences in a large range of
temperatures. Stochastic calculations are shown with black points.
The standard results (Eq. (\ref{thermo})) are given with solid
lines. The parameters of the corresponding dynamics are:
$\epsilon=1.5$, $\delta=1.0$ and $\gamma=0.01$.}
\end{figure}

\begin{figure}[h!]
\includegraphics[width=0.45\textwidth,height=0.45\textwidth,angle=-90]
{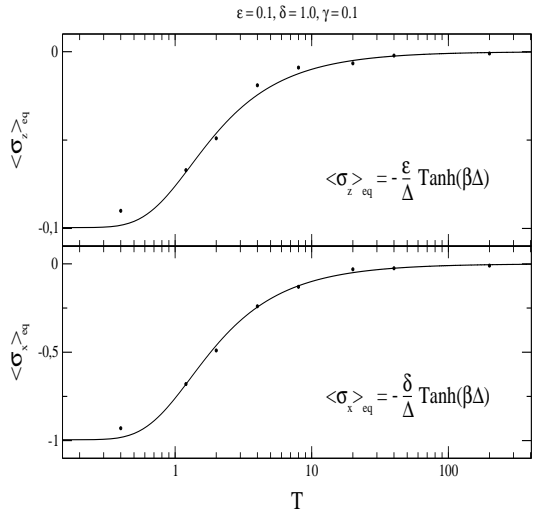} \caption{\label{fig4} Thermal dependence of the
population differences and coherences in a large range of
temperatures. Stochastic calculations are shown with black points.
The standard results (Eq. (\ref{thermo})) are given with solid
lines. The parameters of the corresponding dynamics are:
$\epsilon=0.1$, $\delta=1.0$ and $\gamma=0.1$.}
\end{figure}

Having in mind this limitation, we note that classical noise
effects properly describe the region where both thermal and
internal effects driving the dynamics of the TLS take place (in
particular, for temperatures close to $T_{c}$). In addition, the
effects of the delocalization/localization (given by the ratio
between $\delta$ and $\epsilon$) are correctly taken into account,
as can be seen in Figs. \ref{fig3} and \ref{fig4}. In these two
figures, the thermodynamical functions given by Eqs.
(\ref{thermo}) are plotted (solid lines) for the two main regimes
studied: $\epsilon > \delta$ (Fig. \ref{fig3}), and $\epsilon <
\delta$ (Fig. \ref{fig4}). Similar results have also been obtained
for the regime $\epsilon \sim \delta$. The black points in both
sets of figures are the time asymptotic values issued from the
stochastic dynamics. As can be seen, the agreement is fairly good
at all temperatures studied. Thus, we can conclude that both the
localized and delocalized regimes are correctly described by the
approach here presented. Clearly, the thermodynamical behavior is
independent on the friction value chosen.
%The most significant deviations between the standard thermodynamical calculation and the canonical stochastic
%approach are found in the $\langle \hat \sigma_{x}\rangle_{\beta}$ coherence calculated in the regime
%where $\epsilon = \delta$, which describes competition between tunneling and localization (see Fig. (\ref{fig5})).
%This could be due to...

%\begin{figure}[h!]
%\includegraphics[width=0.45\textwidth,height=0.45\textwidth,angle=-90]
%{fig4.ps} \caption{\label{fig4}
%Thermal dependence of the population differences and coherences in a large
%range of temperatures. Stochastic calculations are shown with dots. The standard
%results (Eq. (\ref{thermo})) are given with solid lines. See text for details.
%}
%\end{figure}

\section{Conclusion}

In this work, a Langevin canonical approach has been developed to
study the thermodynamics of chiral two--level systems interacting
with a thermal bath. Thermal effects due to a noisy environment,
modelled as a collection of classical harmonic oscillators coupled
linearly with the system via a Caldeira--Leggett--like Hamiltonian
have been taken into account. The canonical variables have been
related to both the population differences and coherences of the
system, showing that complete information about the dynamics can
be encoded in them. The competing localization/delocalization
process which is governed by the tunneling and the asymmetry of
the TLS has been analyzed in terms of the critical temperature for
an Ohmic environment. In particular, we have shown that this
critical temperature separates the incoherent and coherent
tunneling regimes. A high number of stochastic trajectories have
been carried out in order to provide time-dependent canonical
thermal averages of the population differences and coherences,
showing that the implementation of the noise term in the dynamics
is properly accounted for. This approach has allowed us to follow
the evolution of the system towards the thermodynamic equilibrium
with fairly good accuracy. Moreover, these stochastic results have
been fitted to analytical expressions beyond the NIBA to extract
more physical information of the parameters ruling this dynamics
such as global damping coefficients as well as Rabi-type
frequencies. The regime of very low temperatures (much smaller
than the critical temperature) which requires a generalized
Langevin equation is not considered.

As has been recently reported by Miller \cite{Bill2012}, an
important topic in particle dynamics is to ascertain the origin of
coherence, classical or quantum, caused by interference of
probability amplitudes. One of the concluding remarks was that
sometimes whether coherence is quantum or classical depends on
what is being observed. When speaking about observation, one has
to think in terms of a measurement apparatus. It is well know that
an environment can be seen as this apparatus \cite{Weiss1999}.
When interacting with the system, it displays decoherence. In our
case, the system always displays tunneling (a typical quantum
feature) but the environment which has been considered to be
classical leads to decoherence. However, coherence at short times
is still observed when the temperature and the friction are small.
Thus, in this context, we deal with the classical observation of
quantum coherence which is destroyed at asymptotic times.

An extension of this work including correlation functions and more
thermodynamics functions such as heat capacity and entropy as well
as the introduction of a magnetic field is currently in progress.

This work has been funded by the MICINN (Spain) through Grant Nos.
CTQ2008-02578, FIS2010-18132, and by the Comunidad Aut\'onoma de
Madrid, Grant No. S-2009/MAT/1467. P. B. acknowledges a Juan de la
Cierva fellowship from the MICINN and A.D.-U. acknowledges a JAE
fellowship from CSIC. H. C. P.-R. and G. R.-L. acknowledge a
scientific project from INSTEC.

\end{document}